\documentclass[aps,prb,twocolumn,amsmath,amssymb,showpacs,groupedaddress]{revtex4}

\usepackage{bm}
\usepackage{graphicx}
\usepackage{color}

\begin{document}


\title{Linear magnetotransport in monolayer MoS$_2$}

\author{C. M. Wang}
\email[]{cmwangsjtu@gmail.com}
\affiliation{School of Physics and
Electrical Engineering, Anyang Normal University, Anyang 455000,
China}
\author{X. L. Lei}
\affiliation{Department of Physics and Astronomy, Shanghai Jiao Tong
University, Shanghai 200240, China\\
Collaborative Innovation Center of Advanced Microstructures, Nanjing University, Nanjing 210093, China}

\date{\today}

\begin{abstract}
 A momentum balance equation is developed to investigate the magnetotransport properties in monolayer molybdenum disulphide 
when a strong perpendicular magnetic field and a weak in-plane electric field are applied simultaneously. 
At low temperature, in the presence of intravalley impurity scattering Shubnikov de Haas oscillation shows up 
accompanying by a beating pattern arising from large spin splitting and its period may halve due to high-order oscillating term 
at large magnetic field for samples with ultrahigh mobility. In the case of intervalley disorders, there exists a magnetic-field 
range where the magnetoresistivity almost vanishes. 
For low-mobility layer, 
a phase-inversion of oscillating peaks is acquired in accordance with recent experiment. At high temperature 
when Shubnikov de Haas oscillation is suppressed, the magnetophonon resonances induced by both optical phonons 
(mainly due to homopolar and Fr\"ohlich modes) and acoustic phonons (mainly due to intravalley transverse and longitudinal acoustic modes) 
emerge for suspended system with high mobility. For the single layer on a substrate, another resonance due to surface optical phonons 
may occur, resulting in a complex behavior of the total magnetoresistance. The beating pattern of magnetophonon resonance due to optical phonons 
can also be observed. However, for nonsuspended layer with low mobility, the magnetoresistance oscillation almost disappears 
and the resistivity increases with field monotonously.
\end{abstract}

\pacs{75.47.-m, 72.20.-i, 81.05.Hd}

\maketitle

\section{Introduction}
The rise of graphene,\cite{novoselov2004electric,geim2007rise} showing outstanding mechanical and electronic properties, 
launched the era of monolayer material. However, pristine graphene does not have a band gap, a property essential 
for electronic applications. Although, it is possible to open small band gap in graphene by some method,\cite{Pedersen2008,Li2007quasi} 
it will inevitably lead to increased fabrication complexity and reduced performance of devices.\cite{li2008chemically,jiao2009narrow} 
This produces great limitations on its becoming a perfect candidate for the next generation nanoelectronic material.  
In contrast to graphene, the transition metal dichalcogenides are semiconductors with naturally occurring band gap, 
which overcomes this problem directly. A prominent representative in this dichalcogenide family is molybdenum disulphide (MoS$_2$). 
Bulk MoS$_2$ has an indirect gap, while monolayer MoS$_2$, which can be isolated by exfoliation techniques similar to graphene, 
is a direct-gap semiconductor with a gap of $1.9\,$eV.\cite{scholz2013plasmons} Due to the large carrier mobility\cite{baugher2013intrinsic}, 
high current carrying capacity\cite{lembke2012breakdown}, strong spin-orbit coupling, and coupling of spin and valley degrees of freedom, 
monolayer MoS$_2$ may become a replacement of graphene or even a candidate for the exploitation of novel valleytronic devices.\cite{xiao2012coupled} 

On the aspect of transport investigation of monolayer MoS$_2$, the linear mobility is close to 200$\,{\rm cm^{2}/Vs}$ at low temperature 
where a high-$\kappa$ gate dielectric was used to suppress the charged-impurity scattering strongly.\cite{radisavljevic2013mobility} 
This value is still lower than the theoretical prediction, where the highest phonon-limited mobility in $n$-type monolayer MoS$_2$ 
is 410$\,{\rm cm^{2}/Vs}$ at room temperature.\cite{kaasbjerg2012phonon} On the other hand, the single layer MoS$_2$ device 
grown by chemical vapor deposition shows low temperature mobility up to 500$\,{\rm cm^{2}/Vs}$, 
where the leading scattering mechanism is believed to be the short-range scatterers at high carrier density.\cite{schmidt2014transport} 
Hence, the main scatterings determining the linear mobility is still open question. Further, looking at all theoretical studies 
on electric transport,\cite{kaasbjerg2012phonon,kaasbjerg2013acoustic,li2013intrinsic} the strong spin-orbit coupling 
in $n$-type monolayer MoS$_2$, which can lead to interesting coupled-spin-valley physics,\cite{xiao2012coupled,feng2012intrinsic,shan2013spin} 
is omitted completely and the energy band is chosen to be a simple parabolic one.

Especially, in magnetotransport the spin-orbit coupling is important, which may result in the beating pattern of Shubnikov de Haas oscillation 
(SdHO)\cite{wang2003mtd} and induce direct magnetoresistance oscillation.\cite{wang2014spin} Due to the spin-valley coupling, 
the magnetic control of the valley degree of freedom in monolayer MoS$_2$ in the presence of normal magnetic field has been 
achieved.\cite{cai2013magnetic} The magneto-optical properties\cite{rose2013spin} and magnetocapacitances\cite{zhou2014fully} 
have been analyzed in this system. However, even the basic SdHO considering all kinds of scattering mechanisms in this single 
layer has not been seriously involved either in theoretical or in experimental works. Only recently, 
the SdHO was observed experimentally for the first time in monolayer and few-layer MoS$_2$.\cite{cui2015multi}
In this paper, we  apply a momentum balance equation to investigate the linear magnetotransport at both low 
and high temperatures including SdHO and magnetophonon resonance (MPR) effect induced by optical and {\it acoustic} phonons for both suspended and nonsuspended samples.

\section{Basic Formulation}
We consider a monolayer of transition metal dichalcogenide MoS$_2$ having large number $N$ carriers in the $x$-$y$ plane. 
These carriers, in addition to interacting with each other, are scattered by the random impurities and coupled with phonons in MoS$_2$ and substrate. 
There exists an external magnetic field $\bm B = (0, 0, B)$ applied along the $z$ direction and a uniform electric field $\bm E$ 
in the monolayer plane. The total Hamiltonian of the system is given by
\begin{equation}\label{}
\mathcal H=\mathcal H_{\rm e}+\mathcal H_{\rm ph}+\mathcal H_{\rm
ei}+\mathcal H_{\rm ep}.
\end{equation}
Here the carrier part $\mathcal H_{\rm e}=\sum_j(h_j+ e\bm r_j\cdot\bm E)+\sum_{i<j}V_c({\bm r}_i-{\bm r}_j)$ with $\bm r_j=(x_j,y_j)$ 
being the in-plane coordinate for the $j$th carrier and $-e$ denoting its charge, 
 and $V_c({\bm r}_i-{\bm r}_j)$ standing for the Coulomb coupling 
potential between the $i$th and $j$th carriers, which, though depends solely on ${\bm r}_i-{\bm r}_j$ for the
thin-layer system, may vary with the spatial ($z$-direction) dielectric environment around the layer. 
The single-carrier low-energy Hamiltonian near $K$ ($-K$) point in the Brillouin zone 
for the $j$th carrier in the presence of magnetic field is given by\cite{xiao2012coupled,shan2013spin}
\begin{equation}
h_j=at(\tau_j\pi_{jx}\sigma_{jx}+\pi_{jy}\sigma_{jy})+\frac{\Delta}{2}\sigma_{jz}-\lambda\tau_j\hat s_{jz}\otimes\frac{\sigma_{jz}-1}{2}.
\end{equation}
Here $a$ is the lattice constant, $t$ is the hopping integral, $\Delta$ is the energy gap, $\lambda$ is the spin-orbit coupling parameter, 
$\tau_j=\pm1$ is the valley index of the $j$th carrier referring to $\pm K$ valley, ${\bm \pi}_j\equiv\bm p_j+e\bm A(\bm r_j)=(\pi_{jx}, 
\pi_{jy})$ is the canonical momentum with $\bm p_j=(p_{jx},p_{jy})$ being the momentum of the $j$th carrier and the vector potential 
in the Landau gauge $\bm A(\bm r_j)=(-By_j,0)$, ${\bm \sigma}_j=(\sigma_{jx},\sigma_{jy},\sigma_{jz})$ is its pseudospin operator 
acting on the orbit $\{d_{z^2} ,(d_{x^2-y^2} + i\tau_jd_{xy})/\sqrt{2}\}$ and $\hat s_{jz}$ is the $z$-component of its real spin 
operator. There are $Q$ valleys locating near the halfway points along the ${\it \Gamma}$--$K$ axes, which may introduce additional intervalley 
scattering process, and thus influence the carrier transport.\cite{li2013intrinsic} However, the accurate value of the energy 
separation between the $K$ and $Q$ points is still unsettled\cite{Cheiwchanchamnangij,kaasbjerg2012phonon,li2013intrinsic} 
with the estimate larger than 50\,meV. In the present calculation, the Fermi energy of this discussed system is far below $Q$ valleys. 
Hence, we can safely neglect the limited effect of $Q$ valleys and assume the system in the $K$-valley dominated carrier transport regime. 
$\mathcal H_{\rm ph}$, $\mathcal H_{\rm ei}$ and $\mathcal H_{\rm ep}$ are phonon Hamiltonian, carrier-impurity and carrier-phonon 
interaction, whose forms can be found in the textbook\cite{Mahan} and Refs.\,\onlinecite{lei1987nonlinear} and \onlinecite{lei2008balance}.

In terms of the center-of-mass (c.m.) momentum and coordinate defined as $\bm P=\sum_{j}\bm p_j$ and $\bm R=N^{-1}\sum_{j}\bm r_j$ 
for the whole system of $N$ carriers 
and the relative-carrier momentum and coordinate $\bm p'_j=\bm p_j-\bm P/N$ and $\bm r_j'=\bm r_j-\bm R$ 
of the $j$th carrier,\cite{lei1985gsf,lei2008balance} the carrier Hamiltonian $\mathcal H_{\rm e}$ of this coupled many-body system 
can be written as the sum of a c.m. part $\mathcal H_{\rm cm}$ and a relative carrier part $\mathcal H_{\rm er}$,
 $\mathcal H_{\rm e}=\mathcal H_{\rm cm}+\mathcal H_{\rm er}$, with
\begin{align}\label{}
 \mathcal H_{\rm cm}=&\frac{1}{N}\sum_jat(\tau_j\varPi_x\sigma_{jx}+\varPi_y\sigma_{jy})+Ne{\bm E}\cdot \bm R,\nonumber\\=&\bm V\cdot{\bm \varPi}+Ne{\bm E}\cdot \bm R,\\
\mathcal H_{\rm er}=&\sum_{j}\bigg[at(\tau_j\pi'_{jx}\sigma_{jx}+\pi'_{jy}\sigma_{jy})+\frac{\Delta}{2}
\sigma_{jz}\nonumber\\&-\lambda\tau_j\hat s_{jz}\otimes\frac{\sigma_{jz}-1}{2}\bigg]+
\sum_{i<j}V_c({\bm r}'_i-{\bm r}'_j).
\end{align}
Here ${\bm \varPi}\equiv \bm P+Ne\bm A(\bm R)=(\varPi_x, \varPi_y)$
 is the c.m. canonical momentum of the total system,
${\bm \pi}_j' \equiv \bm p_j'+e\bm A(\bm r_j')=(\pi'_{jx}, \pi'_{jy})$
is the canonical momentum for the $j$th relative carrier, and 
\begin{equation}
\bm V=\dot{\bm R}=-i[{\bm R},{\mathcal H}]=\frac{1}{N}\sum_jat(\tau_j\sigma_{jx}\hat i+\sigma_{jy}\hat j)
\end{equation}
is the c.m. velocity operator of the carrier system.

Note that, the commutation relation between the  c.m. part $\mathcal H_{\rm cm}$ and the relative-carrier part $\mathcal H_{\rm er}$ 
is of order of $1/N$. Hence for a macroscopically large $N$ system the c.m. motion and the relative motion of carriers are 
truly separated from each other. A spatially uniform electric field $\bm E$ 
shows up only in the c.m. part $\mathcal H_{\rm cm}$, and $\mathcal H_{\rm er}$ is just the Hamiltonian of a monolayer MoS$_2$ 
subject to a perpendicular magnetic field {\it without the electric field}. The coupling of two parts appears only 
through the carrier-impurity and carrier-phonon interactions.

To proceed the calculation of transport properties in monolayer MoS$_2$ in the presence of a magnetic field, we can write down 
all the physical quantities in the Landau representation. The Landau levels of the single-particle Hamiltonian $h$ is labeled 
by a band index $\alpha=\pm1$ for conduction and valence band, valley index $\tau=\pm1$ for $K$ and $-K$ valley, and spin index $s=\pm1$ 
for spin up and spin down in addition to the Landau index $n$ with the form
\begin{equation}\label{leveln}
\varepsilon_{\alpha\tau ns}=\tau s\bar\lambda+\alpha\sqrt{\left({\bar\Delta}-\tau s{\bar\lambda}\right)^2+n\omega_c^2},
\end{equation}
for $n=1,2,3\cdots$, while for $n=0$
\begin{equation}\label{level0}
\varepsilon_{\tau 0 s}=-\tau\left({\bar\Delta}-s{\bar\lambda}\right)+s{\bar\lambda},
\end{equation}
with $\bar\Delta=\Delta/2$, $\bar \lambda=\lambda/2$, and the cyclotron frequency $\omega_c=\sqrt{2}at/l_{\rm B}=\sqrt{2|e|B}at$. 
One should take notice of the fact that the zero level ($n=0$) for $K$ valley ($\tau=+1$) is in the valence band, 
while the zero level ($n=0$) for $-K$ valley ($\tau=-1$) is in the conduction band. 
The corresponding eigenstates, including zero levels ($n=0$), are expressed as 
$\Psi_{\alpha\tau ns}=\chi_s\otimes\varphi_{n,s}^{\alpha,\tau}(\bm r,k_x)$,  
with $\chi_s$ standing for the eigenstate of $\hat s_z$ and 
\begin{equation}\label{}
\varphi^{\alpha,+1}_{n,s}(\bm r,k_x)=\frac{e^{ik_xx}}{\sqrt{\Theta^{\alpha,+1}_{n,s}}}\left(
                                                    \begin{array}{c}
                                                      \Lambda^{\alpha,+1}_{ n,s}\phi_{{n-1},k_x}(y) \\
                                                      \phi_{n, k_x}(y) \\
                                                    \end{array}
                                                  \right),
\end{equation}
\begin{equation}\label{}
\varphi^{\alpha,-1}_{n,s}(\bm r,k_x)=\frac{e^{ik_xx}}{\sqrt{\Theta^{\alpha,-1}_{n,s}}}\left(
                                                    \begin{array}{c}
                                                      \phi_{{n},k_x}(y) \\
                                                      \Lambda^{\alpha,-1}_{n,s}\phi_{{n-1},k_x}(y) \\
                                                    \end{array}
                                                  \right).
\end{equation}
Here $k_x$ is the $x$-component of wave vector $\bm k$, the coefficient
\begin{equation}
\Lambda^{\alpha,\tau}_{n,s}=\frac{\sqrt{n}\omega_c}{(\bar \Delta-\tau s \bar\lambda)-\alpha\tau\sqrt{(\bar \Delta-\tau s \bar\lambda)^2+n\omega_c^2}},
\end{equation}
and $\Theta^{\alpha,\tau}_{n,s}=(\Lambda^{\alpha,\tau}_{n,s})^2+1$. Note that for $K$ valley $\tau=+1$ ($-K$ valley $\tau=-1$), 
only the valence band $\alpha=-1$ (conduction band $\alpha=+1$) is allowed when $n=0$. $\phi_{n,k_x}(y)$ is the harmonic oscillator eigenfunction giving by
\begin{equation}\label{}
\phi_{n,k_x}(y)=\frac{1}{\sqrt{2^nn!l_{\rm
B}\sqrt{\pi}}}\exp\left[-\frac{(y-y_c)^2}{2l_{\rm
B}^2}\right]H_n\left(\frac{y-y_c}{l_{\rm B}}\right),
\end{equation}
with $H_n(x)$ the Hermite polynomial, and $y_c=k_x/(eB)$. In the Landau representation, the carrier-impurity and carrier-phonon Hamiltonians 
including both intravalley and intervalley interactions have the following forms:
\begin{align}
\mathcal H_{\rm ei}=&\sum_{\bm q,a}\sum_{\substack{\alpha,\tau,n,s\\\alpha',\tau',n',s'}}U_{\tau\tau'}(\bm q)J_{\alpha\tau ns}^{\alpha'\tau'n's'}(\bm q)e^{i\bm q\cdot(\bm R-\bm r_a)}\nonumber\\&\times c^\dag_{\alpha\tau ns}c_{\alpha'\tau'n's'},\\
\mathcal H_{\rm ep}=&\sum_{\bm q,\nu}\sum_{\substack{\alpha,\tau,n,s\\\alpha',\tau',n',s'}}M_{\tau\tau'}(\bm q,\nu)J_{\alpha\tau ns}^{\alpha'\tau'n's'}(\bm q)\phi_{\bm q\nu}e^{i\bm q\cdot\bm R}\nonumber\\&\times c^\dag_{\alpha\tau ns}c_{\alpha'\tau'n's'}.
\end{align}
Here $U_{\tau\tau'}(\bm q)$ and $M_{\tau\tau'}(\bm q,\nu)$ are the intravalley or intervalley carrier-impurity scattering potential 
with $\bm r_a$ being the impurity position and carrier-phonon coupling matrix of $\nu$ branch, respectively; $c_{\alpha\tau ns}$ 
and $c^\dag_{\alpha\tau ns}$ are the annihilation and creation operators of carrier; $\phi_{\bm q\nu}=b_{\bm q\nu}+b_{-\bm q\nu}^\dag$ 
is the phonon field operator with $b_{\bm q\nu}$ and $b_{\bm q\nu}^\dag$ being the annihilation and creation operators for a two-dimensional (2D) phonon 
of wave vector $\bm q$ in the branch $\nu$ having frequency $\Omega_{\bm q\nu}$; and the integral
\begin{equation}
J_{\alpha\tau ns}^{\alpha'\tau'n's'}(\bm q)=\int {d\bm r'\left\langle\varphi^{\alpha,\tau}_{n,s}(\bm r',k_x)\left|e^{i\bm q\cdot\bm r'}\right|\varphi^{\alpha',\tau'}_{n',s'}(\bm r',k_x)\right\rangle}.
\end{equation}

The derivation of momentum balance equation starts from the rate of change of the c.m. canonical momentum $\dot{\bm \varPi}=-i[\bm \varPi,\mathcal H]$. 
To linear order in the carrier-impurity and carrier-phonon couplings,\cite{lei1985gsf,lei1985tdb,lei2008balance} the statistical average 
of this operator equation can be obtained by using the initial density matrix 
$\hat \rho_0=Z^{-1}e^{-(\mathcal H_{\rm ph}+\mathcal H_{\rm er})/T}$ at temperature $T$ in the case of weak in-plane electric field ${\bm E}$. 
In the dc steady state, $\langle\dot{\bm \varPi}\rangle=0$, the momentum balance equation for a system of unit area 
($N$ is thus understood as the carrier number density) reads
\begin{align}\label{}
0=&-Ne\bm v\times\bm B-Ne\bm E+\bm f_{\rm ei}+\bm f_{\rm ep},\label{forceEq}
\end{align}
with $\bm v=\langle\bm V\rangle$ being the averaged carrier drift velocity. The frictional forces experienced by the center of
mass due to impurity and phonon scatterings, ${\bm f}_{\rm ei}$ and ${\bm f}_{\rm ep}$, have the following form:
\begin{align}\label{}
\bm f_{\rm ei}=&\,\,n_{\rm i}\!\sum_{\bm q,\tau,\tau'}\left|U_{\tau\tau'}(\bm q)\right|^2\bm q\Pi_2^{\tau\tau'}(\bm q,\omega_0),\label{fim}\\
\bm f_{\rm ep}=&\sum_{\bm q,\tau,\tau',\nu}\left|M_{\tau\tau'}(\bm q,\nu)\right|^2\bm q\Pi_2^{\tau\tau'}(\bm q,{\it \Omega}_{\bm q\nu}+\omega_0)\nonumber\\
&\hspace{0.4cm}\times\left[n\Big(\frac{{\it \Omega}_{\bm q\nu}}{T}\Big)-n\Big(\frac{{\it \Omega}_{\bm q\nu}+\omega_0}{T}\Big)\right].\label{fph}
\end{align}
In the above expressions, $n_{\rm i}$ is an effective impurity density; $n(x)=(e^x-1)^{-1}$ is the Bose distribution function; 
$\omega_0\equiv\bm q\cdot\bm v$; $\Pi_2^{\tau\tau'}(\bm q,\omega)$ is the imaginary part of the Fourier spectrum of 
the valley-dependent relative-carrier density correlation function, defined by
\begin{equation}
\Pi^{\tau\tau'}(\bm q,t-t')=-i\theta(t-t')\left\langle\left[\rho_{\bm q}^{\tau\tau'}(t),\rho_{-\bm q}^{\tau'\tau}(t')\right]\right\rangle_0,
\end{equation}
where $\rho_{\bm q}^{\tau\tau'}(t)=e^{i\mathcal H_{\rm er}t}\rho_{\bm q}^{\tau\tau'}e^{-i\mathcal H_{\rm er}t}$ with 
$$\rho_{\bm q}^{\tau\tau'}=\sum_{\substack{\alpha,n,s\\\alpha',n',s'}}J_{\alpha\tau ns}^{\alpha'\tau'n's'}(\bm q)c^\dag_{\alpha\tau ns}c_{\alpha'\tau'n's'},$$ 
and $\langle\cdots\rangle_0$ stands for the statistical averaging with respect to the initial density matrix $\hat \rho_0$.\cite{lei2008balance,lei1985gsf}

 In most cases the electron density-correlation function in the presence of intercarrier coupling, $\Pi_2^{\tau\tau'}(\bm q,\omega)$,
can be obtained in the random-phase approximation through the density-correlation function  $\Pi_{02} ^{\tau\tau'} (\bm q,\omega)$
in the absence of intercarrier coupling,
\begin{equation} 
\Pi_2^{\tau\tau'}(\bm q,\omega)=\frac{\Pi_{02}^{\tau\tau'}(\bm q,\omega)}{|\varepsilon_{\tau\tau'}({\bm q},\omega)|^2},
\end{equation}
where $\varepsilon_{\tau\tau'}({\bm q},\omega)$ is the carrier-coupling related RPA screening function or carrier screening function, which may
vary with the dielectric environment of two-dimensional (2D) monolayer. 
Therefore, in Eqs.(\ref{fim}) and (\ref{fph}) $\Pi_2^{\tau\tau'}(\bm q,\omega)$ function can be replaced by $\Pi_{02}^{\tau\tau'}(\bm q,\omega)$
function, as long as the impurity and phonon scattering potentials are considered screened by the intercarrier coupling:
$U_{\tau\tau'}(\bm q)/\varepsilon_{\tau\tau'}({\bm q},\omega)$ and 
$M_{\tau\tau'}(\bm q,\nu)/\varepsilon_{\tau\tau'}({\bm q},\omega)$. 

The $\Pi_{02}^{\tau\tau'}(\bm q,\omega)$ function can be expressed as
\begin{align}
\Pi_{02}^{\tau\tau'}(\bm q,\omega)=&\frac{1}{2\pi l_{\rm B}^2}\sum_{\substack{\alpha,n,s\\\alpha',n',s'}}
C_{\alpha\tau ns}^{\alpha'\tau'n's'}(z)\nonumber\\&\times\Pi_{02}^{\tau\tau'}(\alpha,n,s;\alpha',n',s';\omega).
\end{align}
Here\cite{ting1977theory}
\begin{align}
&\Pi_{02}^{\tau\tau'}(\alpha,n,s;\alpha',n',s';\omega)=-\frac{1}{\pi}\int_{-\infty}^{+\infty} d\epsilon [f(\epsilon)-f(\epsilon+\omega)]\nonumber\\
&\hspace{2.cm}\times{\rm Im}G_{\alpha\tau ns}(\epsilon+\omega){\rm Im}G_{\alpha'\tau'n's'}(\epsilon), \label{Imgg}
\end{align}
with ${\rm Im}G_{\alpha\tau ns}(\epsilon)$ standing for the imaginary part of retarded Green's function $G_{\alpha\tau ns}(\epsilon)$ 
and the form factor for intravalley case is given by
\begin{align}
&C_{\alpha\tau ns}^{\alpha'\tau n's'}(z)=\delta_{ss'}\frac{1}{\Theta^{\alpha,\tau}_{n,s}\Theta^{\alpha',\tau}_{n',s'}}z^{n_2-n_1}e^{-z}\frac{n_1!}{n_2!}\nonumber\\&\hspace{0.5cm}\times\left[\Lambda^{\alpha,\tau}_{n,s}\Lambda^{\alpha',\tau}_{n',s'}\sqrt{\frac{n_2}{n_1}}L_{n_1-1}^{n_2-n_1}(z)+L_{n_1}^{n_2-n_1}(z)\right]^2,
\end{align}
while for intervalley case it has a more complex form
\begin{widetext}
\begin{align}
C_{\alpha\tau ns}^{\alpha'\bar\tau n's'}&(z)=\delta_{ss'}\frac{1}{\Theta^{\alpha,\tau}_{n,s}\Theta^{\alpha',\bar\tau}_{n',s'}}e^{-z}\bigg\{\left(\Lambda^{\alpha,\tau}_{n,s}\right)^2z^{m_2-m_1}\frac{m_1!}{m_2!}\left[L_{m_1}^{m_2-m_1}(z)\right]^2+\left(\Lambda^{\alpha',\bar\tau}_{n',s'}\right)^2z^{k_2-k_1}\frac{k_1!}{k_2!}\left[L_{k_1}^{k_2-k_1}(z)\right]^2\nonumber\\&+2s_1^{m_2-m_1}s_2^{k_2-k_1}\Lambda^{\alpha,\tau}_{n,s}\Lambda^{\alpha',\bar\tau}_{n',s'}z^{(m_2-m_1+k_2-k_1)/2}L_{m_1}^{m_2-m_1}(z)L_{k_1}^{k_2-k_1}(z)\cos[s_1(m_2-m_1)-s_2(k_2-k_1)]\theta_{\bm q}\bigg\},
\end{align}
\end{widetext}
with $L_n^m(z)$ being associated Laguerre polynomials, $z=l_{\rm B}^2q^2/2$, $n_1=\min(n,n')$, $n_2=\max(n,n')$, $m_1=\min(n-1,n')$, $m_2=\max(n-1,n')$, $k_1=\min(n,n'-1)$, $k_2=\max(n,n'-1)$, $\theta_{\bm q}$ is the polar angle of wave vector $\bm q$, and
\begin{equation*}
s_1=\left\{\begin{array}{cr}
1,&\,\,n-1<n'\\
-1,&\,\,n-1\ge n'
                \end{array}\right.,
\end{equation*}
\begin{equation*}
s_2=\left\{\begin{array}{cr}
1,&\,\,n<n'-1\\
-1,&\,\,n\ge n'-1
                \end{array}\right..
\end{equation*}

In the presence of carrier-impurity, carrier-phonon, and carrier-carrier scatterings, the Landau levels of monolayer MoS$_2$ are broadened. The imaginary part of the retarded Green's function ${\rm Im}G_{\alpha\tau ns}(\epsilon)$ or the density of state of the $\alpha\tau ns$th Landau level is modeled using a Gaussian form:\cite{Ando1982}
\begin{equation}\label{}
{\rm Im}G_{\alpha\tau ns}(\epsilon)=-\frac{\sqrt{2\pi}}{\varGamma_{\alpha\tau ns}}
\exp\left[-\frac{2(\epsilon-\varepsilon_{\alpha\tau ns})^2}
{\varGamma_{\alpha\tau ns}^2}\right],
\end{equation}
with $\varGamma_{\alpha\tau ns}$ denoting the half width.

The chemical potential $\varepsilon_f$ at temperature $T$ is determined by the carrier density (electron density $N_+$ or hole density $N_-$) of the system by the following equation
\begin{align}\label{density}
\left\{
\begin{array}{c}
N_+\\N_-
\end{array}
\right\}=-\frac{1}{2\pi^2l_{\rm B}^2}\sum_{\tau,n,s}\int_{-\infty}^{+\infty}d\epsilon
\left\{
\begin{array}{c}
f(\epsilon){\rm Im}G_{+\tau ns}(\epsilon)\\ 
\left[1-f(\epsilon)\right]{\rm Im}G_{-\tau ns}(\epsilon)\end{array}
\right\}.
\end{align}
Here $f(\epsilon)=\{\exp[(\epsilon-\varepsilon_{f})/T]+1\}^{-1}$ is the Fermi distribution function. For electron conduction case, 
the summation index $n$ in the above equation is taken over $1,2,3,\cdots$ for $K$ valley ($\tau=+1$), but
$0,1,2,\cdots$ for $-K$ valley ($\tau=-1$). However, for hole conduction, it is taken over $0,1,2,\cdots$ for $K$ valley, 
but $1,2,3,\cdots$ for $-K$ valley.

The momentum balance equation \eqref{forceEq} combining with equation of carrier density \eqref{density} describes the steady-state magnetotransport 
of monolayer MoS$_2$, which can determine either the drift velocity (charge current density) for given electric field or 
the electric field for given current.  
In the Hall configuration, e.g., with the charge current $\bm J$ (or drift velocity) in the $x$ direction, 
$\bm J=(J,0)=(-Nev,0)$, the momentum balance equation (\ref{forceEq}) gives a transverse magnetoresistance $R_{xy}=-E_y/(Nev)=-B/(Ne)$ 
and a longitudinal magnetoresistance $R_{xx}=-(f_{\rm ei}+f_{\rm ep})/(N^2e^2v)$.

\section{numerical results and discussion}\label{num}
For numerical calculation, we concentrate on the $n$-doped case, i.e., carrier is electron, $N=N_+$ and we only need to consider 
$\alpha=\alpha'=+1$. In the following, the index $\alpha$ or $\alpha'$ will be omitted. 
The half-width $\varGamma_{\tau ns}$ 
should vary with the band indices generally. However, for simplicity, we neglect the effect of spin-orbit interaction, 
and take it with the form:\cite{lei2003radiation,lei2005magn} 
\begin{equation}
\varGamma=\sqrt{\frac{e\omega_{c0}\alpha_{\Gamma}}{\pi m^*\mu}}.
\end{equation}
Here $\mu$ is the zero-field mobility at temperature $T$, $\omega_{c0}=eB/m^*$ is the cyclotron frequency with effective 
mass $m^*=\Delta/(a^2t^2)$, and $\alpha_{\Gamma}$ is a phenomenological parameter to relate the single-particle lifetime to 
the transport scattering time.\cite{lei2003radiation,lei2005magn} In the following numerical evaluation, we will set $\alpha_\Gamma=3$, except otherwise specified.

The intravalley electron-impurity scattering potential is considered 
due to charged impurities distributed at a distance $d$ from the layer:\cite{wang2013nonlinearm}
\begin{equation}
U_{\tau\tau}(\bm q)=\frac{Z_ie^2}{2\varepsilon_0\kappa q}e^{-qd},
\end{equation}
with $Z_i$ standing for the effective impurity charge number,
 $\kappa$ for the dielectric constant of MoS$_2$. For suspended MoS$_2$ monolayer, $d=0$. 
 The carrier screening is taken into account with 
 a static screening function of Thomas-Fermi form\cite{Ando1982,kaasbjerg2013acoustic,Macharge2014}
\begin{equation}
\varepsilon(q,0)=\varepsilon(q)=1+\frac{q_{\rm TF}^{\rm eff}}{q}.
\end{equation}
For suspended MoS$_2$, $q_{\rm TF}^{\rm eff}$ equals the zero-temperature Thomas-Fermi wave vector 
$q_{\rm TF}=m^*e^2/(\pi\varepsilon_0\kappa)$; while for the layer on a substrate, the value of $q_{\rm TF}^{\rm eff}$, 
which could be a couple of times larger or smaller than $q_{\rm TF}$ depending on the dielectric environment and carrier density, 
will be taken from Ref.\,\onlinecite{Macharge2014}. 
Note that the main role of carrier screening is to enhance or decrease the mobility with or without magnetic field.
In the present study, the effective impurity charge density $n_i Z_i^2$, which may be modified by the spatial dielectric environment 
of the system, is determined by the zero-temperature carrier mobility $\mu_0$ in the absence of the magnetic field under the same screening
condition, thus the major magnetic-field related behaviors are not sensitive to the detailed form of the scattering potential or screening. 

 In addition to above intravalley impurity scattering, we also include intervalley disorder scattering, which 
 can be induced by lattice vacancy in a two-dimensional honeycomb lattice\cite{suzuura2002crossover} and by defects raised 
 from ion irradiation as in graphene.\cite{chen2009defect} The scattering potential is usually modeled by a $\delta$-function form, i.e., 
a constant $U_{\tau\bar\tau}(\bm q)=u_0$. 

For intrinsic electron-phonon couplings in suspended layer, we consider both intravalley and intervalley acoustic deformation potential
interactions. In the case of optical deformation potential, both zero-order and first-order couplings are taken into account and 
the homopolar mode is also included. The relevant formulas can be found in Ref.\,\onlinecite{kaasbjerg2012phonon}. 
The polar longitudinal optical phonons are also important and their coupling matrix element with 2D carriers can be written
as\cite{kaasbjerg2012phonon,kaasbjerg2014hot}
\begin{equation}
M_{\tau\tau}(\bm q,{\rm Fr})=g_{\rm Fr}{\rm erfc}(q\sigma/2),
\end{equation}
where $g_{\rm Fr}$ is the Fr\"ohlich coupling constant, erfc is the complementary error function, and $\sigma$ is the effective width 
of the electronic envelope function.

For MoS$_2$ on a substrate, the surface optical phonons (SOPs) couple to the electrons via an effective electric field, 
which may play important role in transport\cite{zeng2013remote,Macharge2014} just like 
graphene.\cite{fratini2008substrate,konar2010effect,wang2013nonlinearm} The coupling matrix element is expressed as\cite{fratini2008substrate}
\begin{equation}
|M_{\tau\tau}(\bm q,{\rm SO})|^2=\frac{e^2\Omega_{\Gamma,\rm so}}{2\varepsilon_0\kappa q}\left(\frac{1}{1+\kappa_e^\infty}-\frac{1}{1+\kappa_e^0}\right)e^{-2qd},
\end{equation}
with $\Omega_{\Gamma,\rm so}$ the frequency of SOP, and $\kappa_e^\infty$ ($\kappa_e^0$) denoting the high (low) frequency dielectric constant 
of substrate.

Unlike the case of the static impurity scattering, the carrier screening for phonon scattering is dynamic, i.e., it is the
 screening function $\varepsilon({\bm q},\Omega_{q\nu})$ at phonon frequency $\Omega_{q\nu}$ rather than the static function 
$\varepsilon({\bm q},0)$, should be used in the equation with phonon scattering. 
It has been shown\cite{lei2008balance,lei1985} that optic (as well as acoustic) phonon induced 2D resistivity with dynamic screening are 
essentially equivalent to those without screening at temperature $T>100$\,K, when phonon scatterings  play
important roles. Therefore, in the numerical calculation of 
phonon-related magnetotransport at higher temperatures  we will not include screening in the electron-phonon matrix element. 

The relevant parameters used in the numerical calculation are listed in Table \ref{tab:para}, except otherwise specified. 

\begin{table}
    \centering
\caption{Material parameters for monolayer MoS$_2$ used for
calculation. The $\bf \Gamma$/${ \bf K}$ subscripts represent intra/intervalley phonons.}
    \label{tab:para}
\begin{ruledtabular}
\begin{tabular}{lll}
  Parameter & Symbol & Value \\
  \hline
  Lattice constant\cite{xiao2012coupled}  & $a$ & $3.193$\,\AA \\
  Hopping integral\cite{xiao2012coupled}  & $t$  & $1.1$\,eV \\
  Energy gap\cite{scholz2013plasmons}     & $\Delta$& $1.9\,{\rm eV}$ \\
  Spin splitting energy\cite{xiao2012coupled} & $\lambda$  & $75\,{\rm meV}$ \\
  Mass density\cite{kaasbjerg2012phonon} & $\rho$  & $3.1\times10^{-7}{\rm g/cm^{2}}$ \\
  Effective Layer thickness\cite{kaasbjerg2012phonon} & $\sigma$  & 4.41\,\AA \\
  dielectric constant of MoS$_2$\cite{kim2012high} & $\kappa$ & 7.6\\
  dielectric constant of ZrO$_2$\cite{konar2010effect} &  & \\
  low frequency & $\kappa_e^0$ & 24\\
  high frequency & $\kappa_e^\infty$ & 4\\
  Transverse sound velocity\cite{kaasbjerg2012phonon} & $v_{\rm TA}$    & $4200\,{\rm m/s}$ \\
  Longitudinal sound velocity\cite{kaasbjerg2012phonon} &  $v_{\rm LA}$     & $6700\,{\rm m/s}$ \\
  Acoustic deformation potentials\cite{kaasbjerg2012phonon} & & \\
  TA & $\Xi_{\rm TA}$ & $1.6\,{\rm eV}$\\
  LA & $\Xi_{\rm LA}$ & $2.8\,{\rm eV}$\\
  TA & $D_{\bf K,\rm TA}^1$ & $5.9\,{\rm eV}$\\
  LA & $D_{\bf K,\rm LA}^1$ & $3.9\,{\rm eV}$\\
  Optical deformation potentials\cite{kaasbjerg2012phonon} & & \\
  TO & $D_{\bf \Gamma,\rm TO}^1$ & $4.0\,{\rm eV}$\\
  TO & $D_{\bf K,\rm TO}^1$ & $1.9\,{\rm eV}$\\
  LO & $D_{\bf K,\rm LO}^0$ & $2.6\times10^8\,{\rm eV/cm}$\\
  Homopolar & $D_{\bf \Gamma,\rm HP}^0$ & $4.1\times10^8\,{\rm eV/cm}$\\
  Fr\"{o}hlich coupling\cite{kaasbjerg2014hot} & &\\
  LO & $g_{\rm Fr}$ & $286\,{\rm meV\AA}$\\
  Phonon energies\cite{li2013intrinsic} & &\\
  TA & $\Omega_{\bf K,{\rm TA}}$& 23.1\,meV\\
  LA & $\Omega_{\bf K,{\rm LA}}$& 29.1\,meV\\
  TO & $\Omega_{\bf \Gamma,{\rm TO}}$& 48.6\,meV\\
  TO & $\Omega_{\bf K,{\rm TO}}$& 46.4\,meV\\
  LO & $\Omega_{\bf \Gamma,{\rm LO}}$& 48.0\,meV\\
  LO & $\Omega_{\bf K,{\rm LO}}$& 42.2\,meV\\
  Homopolar & $\Omega_{\bf \Gamma,{\rm HP}}$& 50.9\,meV\\
  SOP energies of ZrO$_2$\cite{konar2010effect} & &\\
  1st mode & $\Omega_{\bf\Gamma,{\rm so}}^{(1)}$& 25.02\,meV\\
  2nd mode & $\Omega_{\bf\Gamma,{\rm so}}^{(2)}$& 70.8\,meV\\
\end{tabular}
\end{ruledtabular}
\end{table}

\subsection{Shubnikov de Haas oscillation}

\begin{figure}
\begin{center}
  \includegraphics[width=0.4\textwidth]{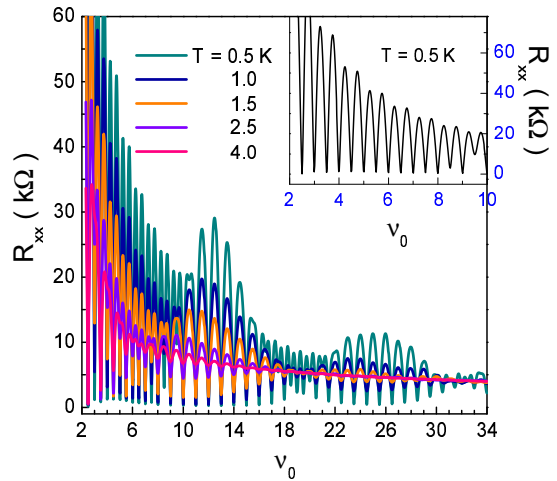}
\end{center}
\caption{(Color online) Longitudinal magnetoresistance as a function of average filling factor $\nu_0$ at different lattice temperatures 
$T=0.5,1.0,1.5,2.5,4.0\,{\rm K}$. The inset shows the enlarged magnetoresistance at $T=0.5\,{\rm K}$ for small filling. 
The linear mobility at zero temperature\cite{li2013intrinsic} $\mu_0=4000\,{\rm cm^2/Vs}$ and the electron density 
$N=7\times10^{12}\,{\rm cm^{-2}}$.}\label{SdHT}
\end{figure}

In this subsection we consider the magnetotransport of suspended MoS$_2$ at low temperatures. First, the impurity scattering is assumed to be only the 
intravalley Coulombic scattering ($d=0$). In Fig.\,\ref{SdHT}, the longitudinal magnetoresistance $R_{xx}$ is calculated versus 
average filling factor $\nu_0=\omega_c^{-2}[\varepsilon_{\rm F}^2-{\bar\Delta}^2]$ with $\varepsilon_{\rm F}$ denoting the Fermi energy. 
The electron density is set to be $N=7\times10^{12}\,{\rm cm}^{-2}$ and the zero-field linear mobility  $\mu_0=4000\,{\rm cm^2/Vs}$ at zero temperature.
This value of mobility, though  
one order larger than those currently obtained experimentally,\cite{radisavljevic2013mobility,schmidt2014transport}
is consistent with the theoretical work.\cite{li2013intrinsic} 
Higher linear mobility at low temperature can be achieved via the gate dielectric engineering to effectively 
screen charge impurities,\cite{Ong2013} and doping and strain modulations already realized a mobility higher than 1000\,cm$^2$/Vs 
at room temperature,\cite{Ge2014} we thus expect this zero-temperature mobility will be reached in the near future. 

As can be seen from Fig.\,\ref{SdHT}, the magnetoresistivity versus filling factor $\nu_0$ or magnetic field $B$, 
exhibits marked SdHO with a beating pattern, having approximate period $\varDelta\nu_0\simeq1$ 
at large fillings or low magnetic fields. The resistivity peaks or valleys locate at integer fillings. 
There is a phase inversion, i.e., a change from the integer fillings for peaks to the ones for valleys. 
These features are in vivid contrast to graphene,\cite{zhang2005experimental,tan2011shubnikov} 
where the SdHO valleys locate in the vicinity of half-integer filling factors without beating patterns, 
but analogous to the behavior of conventional 2D electron gas with spin-orbit coupling.\cite{wang2003mtd} 
Further, at large magnetic fields or small fillings, the period of oscillation in monolayer MoS$_2$ halves, 
which can be seen clearly in the inset of Fig.\,\ref{SdHT}. With an increase of temperature, the amplitude of SdHO decreases rapidly.

\begin{figure}
\begin{center}
  \includegraphics[width=0.4\textwidth]{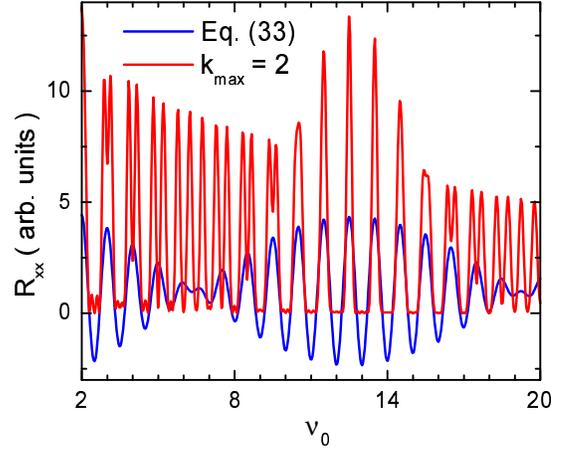}
\end{center}
\caption{(Color online) Curves of analytical expressions for magnetoresistance versus average filling factor $\nu_0$. 
The blue solid line is obtained from Eq. \eqref{keq1}, while the red solid line is directly calculated from Eq.\,\eqref{kinfty} 
for the maximum $k_{\rm max}=2$. The other parameters are the same as Fig.\,\ref{SdHT}.}\label{analy}
\end{figure}

At these low temperatures shown, contribution to the frictional
force mainly originates from electron-impurity scattering, and thus resistivity $R_{xx}\simeq-f_{\rm ei}/(N^2e^2v)$. 
For $\delta$-form intravalley short-range scattering potential $U_{\tau\tau}(\bm q)=u_0$,
the zero-temperature magnetoresistivity $R_{xx}$ can be expressed as 
\begin{align*}
R_{xx}&=-\frac{n_{\rm i}u_0^2}{N^2e^2}\sum_{\bm q\tau}q_x^2\left.\frac{\partial\Pi_{02}^{\tau\tau}(\bm q,\omega)}
{\partial\omega}\right|_{\omega=0}\\&=\frac{n_{\rm i}\pi u_0^2}{N^2e^2l_{\rm B}^2}\sum_{\tau ss'}g_{\tau s}(\varepsilon_{\rm F})
g_{\tau s'}(\varepsilon_{\rm F})\left[\int_0^{+\infty}dz zC_{\tau \nu_{\tau s}s}^{\tau \nu_{\tau s'}s'}(z)\right],
\end{align*}
in which the density of states of electrons in the $\tau$th valley with spin $s$ at Fermi energy $\varepsilon_{\rm F}$,
 $g_{\tau s}(\varepsilon_{\rm F})=-\sum_{n}{\rm Im}G_{\tau n s}(\varepsilon_{\rm F})/(2\pi^2 l_{\rm B}^2)$,
 can be rewritten, by means of Poisson summation formula, as 
\begin{align}
  g_{\tau s}(\varepsilon_{\rm F})=&\frac{\varepsilon_{\rm F}}{\pi l_{\rm B}^2\omega_c^2}\Bigg\{1+2\sum_{k=1}^\infty\big[\cos(2\pi k\nu_{\tau s})\nonumber\\&-k\beta\sin(2\pi k\nu_{\tau s})\big]\exp\left(-2k^2\frac{\beta^2\varepsilon_{\rm F}^2}{\varGamma^2}\right)\Bigg\},
\end{align}
where $\nu_{\tau s}=\omega_c^{-2}(\varepsilon_{\rm F}-\bar\Delta)(\varepsilon_{\rm F}+\bar\Delta-\tau s\bar\lambda)$
is the filling factor of electrons in $\tau$th valley with spin $s$, and $\beta=\pi\varGamma^2/\omega_c^2$. 
For monolayer MoS$_2$ even with low mobility $\mu\sim10\,{\rm cm^2/Vs}$, the coefficient $\beta\ll1$, therefore, 
the term with sine function could be omitted safely.
In the case of high filling factor $\nu_{\tau s}$, the integral
\[\int_0^{+\infty}dz zC_{\tau \nu_{\tau s}s}^{\tau \nu_{\tau s'}s'}(z)=2\nu_{\tau s}\delta_{ss'},\]
and the linear magnetoresistance can be written as
\begin{align}\label{kinfty}
  R_{xx}=&\frac{n_{\rm i}u_0^2}{2\pi N^2e^2}\frac{\varepsilon_{\rm F}^2}{ l_{\rm B}^2a^4t^4}\sum_{\tau s}\nu_{\tau s}\Bigg[1+2\sum_{k=1}^{k_{\rm max}=\infty}\cos(2\pi k\nu_{\tau s})\nonumber\\&\times\exp\left(-2k^2\frac{\beta^2\varepsilon_{\rm F}^2}{\varGamma^2}\right)\Bigg]^2.
\end{align}
Usually, on the account of the rapid decay of the exponential function, one only needs to keep terms with $k=1$ in the summation, leading to 
\begin{align}\label{keq1}
  R_{xx}=&\frac{n_{\rm i}u_0^2}{N^2e^2}\frac{\varepsilon_{\rm F}^2(\varepsilon_{\rm F}^2-\bar\Delta^2)}{\pi a^6t^6}\Bigg[1+4\cos\left(2\pi\nu_0\right)\nonumber\\&\times\cos\left(2\pi\nu_0\frac{\lambda}{\varepsilon_{\rm F}+\bar\Delta}\right)\exp\left(-2\frac{\beta^2\varepsilon_{\rm F}^2}{\varGamma^2}\right)\Bigg].
\end{align}
This represents that the amplitude of oscillation is modulated by the second cosine function due to the spin-splitting and 
there are nodes at ${\lambda\nu_0}/({\varepsilon_{\rm F}+\bar\Delta})=l\pm{1}/{4}$ with $l$ being an integer. 
Note that three smallest nodes in positive regime corresponds to ${\lambda\nu_0}/({\varepsilon_{\rm F}+\bar\Delta})=0.25,0.75,1.25$ 
or $\nu_0=6.4,19.1,31.9$, in agreement with the numerical calculation (see Fig.\,\ref{SdHT}). 
However, the oscillating peaks at large magnetic field or small filling factor obey $\varDelta(\nu_0)\simeq0.5$ in the figure, 
which cannot be explained by the above equation and is due to terms of higher frequency. Because of the small value of $\beta$, 
the product $(\beta\varepsilon_{\rm F}/\varGamma)^2$ may not be considerably larger than one and the oscillating terms with $k>1$ 
also may somewhat contribute to the total resistivity. Fig.\,\ref{analy} demonstrates the results from 
the approximate expression \eqref{keq1} and from \eqref{kinfty} with $k$ summing up to 2. 
It is clear that the oscillation part of high frequency comes from the terms with $k=2$. 
It is noteworthy that this feature is irrespective of the half-integer filling in graphene 
due to the electron-hole symmetry of zero Landau level for massless electrons. In the absence of magnetic field, 
the resistivity $R_{xx}$ reduces to
\begin{equation}
R_0=\frac{n_{\rm i}u_0^2}{N^2e^2}\frac{\varepsilon_{\rm F}^2(\varepsilon_{\rm F}^2-\bar\Delta^2)}{\pi a^6t^6}=\frac{n_{\rm i}u_0^2}{Ne^2}\frac{\pi a^2t^2N+{\bar\Delta}^2}{a^4t^4}.
\end{equation}
Despite the linear dispersion on momentum, this resistivity depends on the electron density owing to its massive property, 
in contrast to the result of graphene.\cite{lineargraphene} The corresponding density $N$ is $58\times10^{12}\,{\rm cm^{-2}}$ 
when $\pi a^2t^2N$ equals ${\bar\Delta}^2$ for the present parameters. Hence, for small density resistivity $R_0$ is inversely 
proportional to density similar to the case of conventional 2D electron gas, while for very large density $R_0$ becomes independent 
of electron density.

\begin{figure}
\begin{center}
  \includegraphics[width=0.45\textwidth]{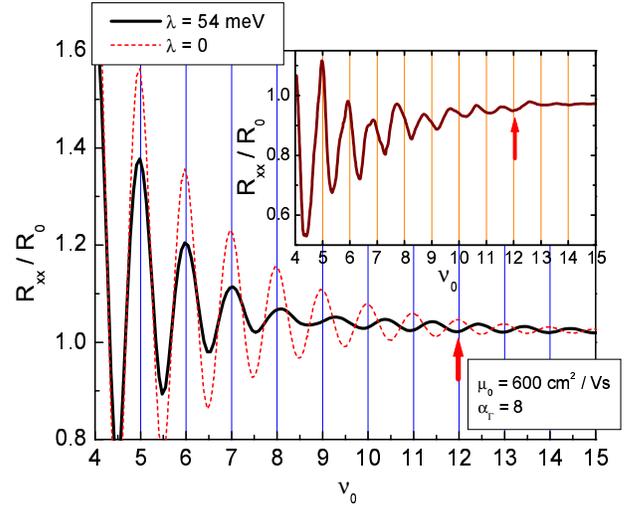}
\end{center}
\caption{(Color online) Normalized magnetoresistance versus filling factor $\nu_0$ for the case $\lambda=54\,{\rm meV}$ (black solid line) 
and $\lambda=0$ (red dash line) at temperature $T=0.3\,{\rm K}$. Here electron density $N=9.69\times10^{12}\,{\rm cm}^{-2}$, 
linear mobility $\mu_0=600\,{\rm cm^2/Vs}$, and $\alpha_\Gamma=8$. The inset shows the corresponding experimental results, 
which are replotted as function of $\nu_0$.}\label{compa}
\end{figure}

\begin{figure}
\begin{center}
  \includegraphics[width=0.4\textwidth]{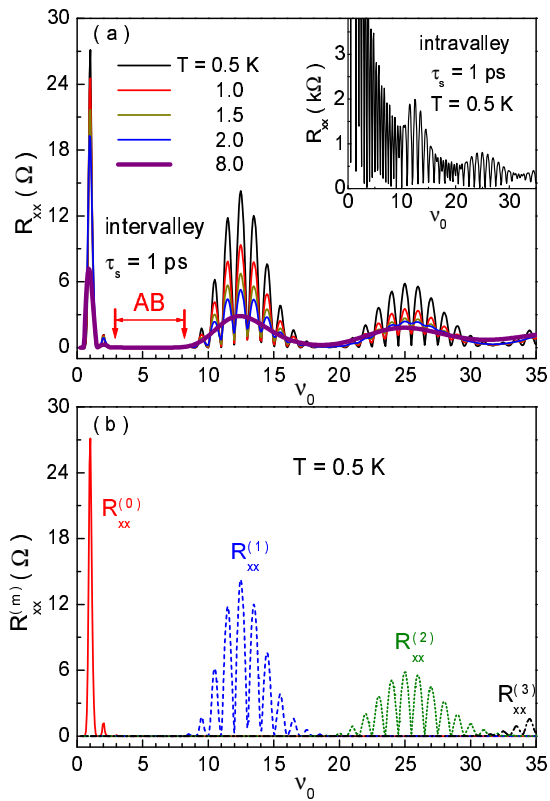}
\end{center}
\caption{(Color online) (a) Intervalley electron-impurity scattering induced magnetoresistance vs the average filling factor $\nu_0$ 
at different lattice temperatures $T=0.5,1.0,1.5,2.0,8.0\,{\rm K}$ when the relaxation time $\tau_s=1\,{\rm ps}$. 
The inset shows the intravalley short-range electron-impurity scattering induced magnetoresistance at $T=0.5\,{\rm K}$ 
for the same relaxation time. 
(b) Magnetoresistance contributions $R_{xx}^{(0)}$, $R_{xx}^{(1)}$, $R_{xx}^{(2)}$ and $R_{xx}^{(3)}$ versus the average filling factor 
at $T=0.5\,{\rm K}$.}\label{inter}
\end{figure}

To compare our theoretical result with recent experimental observation,\cite{{cui2015multi}} 
Fig.\,\ref{compa} presents the normalized resistivity versus filling factor for another monolayer MoS$_2$  
with electron density $N=9.69\times10^{12}\,{\rm cm}^{-2}$ and linear mobility $\mu_0=600\,{\rm cm^2/Vs}$ at $T=0.3\,{\rm K}$. 
The curves for the cases with spin-orbit coupling $\lambda=54\,{\rm meV}$ and without spin-orbit coupling are plotted, respectively, 
as solid and dash lines. Here the factor $\alpha_\Gamma=8$. In the inset, we replot the experimental result taken from 
Fig.\,4(b) in Ref.\,\onlinecite{cui2015multi}, as a function of average filling factor. For the experimental sample having low mobility, 
the Landau level broadening is so large that the beating pattern can not be observed. Nevertheless, the phase inversion of SdHO peaks 
still shows clearly. As can be seen, $\nu_0=5$ corresponds to a position of SdHO peak, while $\nu_0=12$ is for valley. 
The numerical calculation agrees with the experimental observation well. The red dash line for the case of $\lambda=0$, 
where the peaks always locate at integer filling factors, is also plotted for comparison. 
We can see that the spin-orbit splitting is very important for magnetotransport in monolayer MoS$_2$, even for low-mobility sample 
in which the full beating pattern of SdHO is not easy to observe.

To investigate the intervalley scattering effect on the SdHO, in Fig.\,\ref{inter} we plot the oscillating magnetoresistance induced 
solely by the short-range intervalley disorder at various lattice temperatures $T=0.5,1.0,1.5,2.0,8.0\,{\rm K}$. 
Here the relaxation time $\tau_s=1/(m^*n_{\rm i}u_0^2)$ is set to be 1\,ps. For the purpose of comparison, the SdHO induced by intravalley 
short-range electron-impurity scattering is also plotted in the inset of Fig.\,\ref{inter}(a) for the same value of relaxation time 
at $T=0.5\,{\rm K}$. It is seen that the magnetoresistance induced by the intervalley collision, though almost two order smaller than 
intravalley one, also exhibits SdHO versus the average filling factor and the extrema show up at integer fillings and 
the oscillation is also modulated by the spin-orbit interaction with nodes locating at the same positions as in the intrasuband case. 
But the modulation appears much stronger than the intravalley one: with increasing temperature the amplitude of SdHO decreases, 
while the envelope of oscillation still exists even at $T=8.0\,$K when the intravalley one disappears. 
Especially, in contrast to the intravalley case, there exists a regime $AB$ ($3<\nu_0<8$ or $9\,{\rm T}<B<24\,{\rm T}$), 
in which the magnetoresistance almost vanishes. 

All these can be referred to the fact that, in contrast to intravalley case, the intervalley scattering hardly takes place 
between two states having the same Landau index $n>0$. The Landau levels $\varepsilon_{\tau n s}$ expressed in (\ref{leveln}) for $n=1,2,3,...$, 
can be written as $\epsilon_{n,\iota}$ with $\iota\equiv \tau s$.
As indicated in Eq.\,(\ref{Imgg}) the resistivity is proportional to the product of DOSs of two close (contributory) Landau levels 
around the Fermi energy. In the vicinity of Fermi energy, for a fixed Landau index $n$ the level separation of different $\iota$ 
is almost independent of the magnetic field, while the distance between Landau levels having same $\iota$ but different 
Landau indexes $n$ and $n'$ is proportional to the magnetic field. 
Hence, at large magnetic fields two contributory Landau levels of different $\iota$ must have the same Landau index $n$. 
At low magnetic fields, the Landau indexes of two contributory Landau levels may not be equal to each other and their difference increases 
with decreasing magnetic field. In Fig.\,\ref{inter}(b) the magnetoresistance $R_{xx}^{(m)}$, contributed from electron transitions 
between two Landau levels with Landau-index difference of $m$ near Fermi energy, are plotted as functions of average filling factor 
$\nu_0$ at 0.5\,K. At low filling factors or large magnetic fields, the energy distance between levels with same Landau and spin indexes 
but different valley indexes is smaller compared with that between Landau levels with different Landau indexes, and 
we only need to consider the transition between levels of different valleys but having same Landau index $\nu_0$, leading to $R_{xx}^{(0)}$. 
For low magnetic fields, contributions of electron transitions between levels having different Landau indexes dominate. 
Here, $R_{xx}^{(1)}$ stands for contribution from the transitions between $\nu_0$ and $\nu_0-1$ levels and those between $\nu_0$ and $\nu_0+1$ 
levels. $R_{xx}^{(2)}$ stands for contribution from electron transitions between $\nu_0+1$ and $\nu_0-1$ levels, 
and $R_{xx}^{(3)}$ for contribution from transitions between $\nu_0+2$ and $\nu_0-1$ levels and those between $\nu_0-2$ and $\nu_0+1$ levels.
It is found that $R_{xx}^{(0)}+R_{xx}^{(1)}+R_{xx}^{(2)}+R_{xx}^{(3)}$ almost equals the total magnetoresistance $R_{xx}$ shown 
in Fig.\,\ref{inter}(a). 

$R_{xx}^{(0)}$ becomes quite small when $\epsilon_{\nu_0,+}-\varepsilon_{\rm F}\gtrsim\Gamma$ 
and/or $\varepsilon_{\rm F}-\epsilon_{\nu_0,-}\gtrsim\Gamma$, i.e., it is almost zero for magnetic fields lower than a certain value.
On the other hand, with the increase of the magnetic field, the level distance of different Landau indexes enlarges,
leading to $R_{xx}^{(1)}$ almost vanishing for magnetic fields larger than a certain value, which is determined by $\epsilon_{\nu_0,-}-\epsilon_{\nu_0-1,+}\gtrsim\Gamma$ (so do for $R_{xx}^{(2)}$ and $R_{xx}^{(3)}$). 
In the range between these two magnetic fields, the total magnetoresistance appears very small. 
For the present parameters (set $\epsilon_{\nu_0,+}-\epsilon_{\rm F}=1.3\Gamma$), this range is $8.8\,{\rm T}<B<23.3\,{\rm T}$ 
or $3.1<\nu_0<8.2$, as indicated $AB$ in the Fig.\,\ref{inter}.

\subsection{Magnetophonon resonance}

\begin{figure}
\begin{center}
  \includegraphics[width=0.48\textwidth]{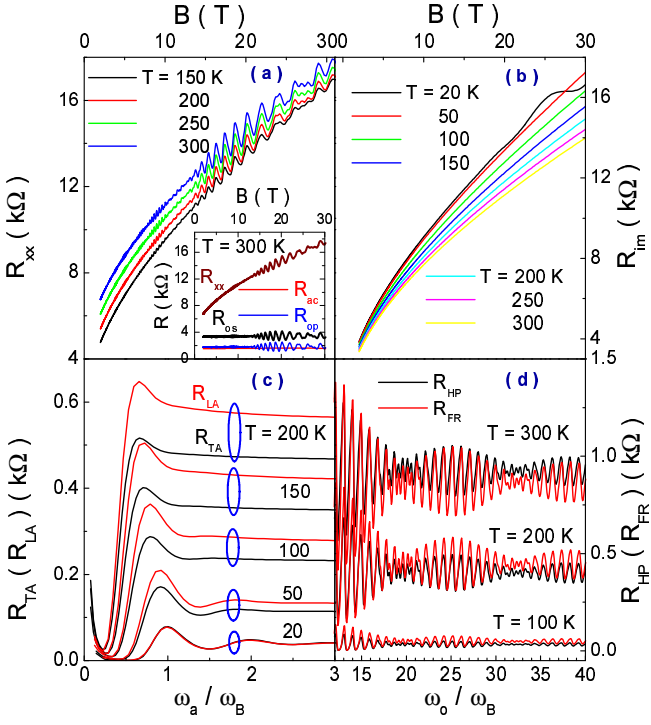}
\end{center}
\caption{(Color online) (a) Total longitudinal magnetoresistance $R_{xx}$ in suspended MoS$_2$ versus magnetic field $B$ 
at high temperatures $T=150,200,250,300\,{\rm K}$. The inset shows the main contributions from electron-phonon scattering, 
where $R_{\rm os}=R_{\rm ac}+R_{\rm op}$. (b) The impurity-induced magnetoresistance $R_{\rm im}$ is plotted as a function of magnetic field $B$. 
(c) The magnetoresistance induced by intravalley transverse (longitudinal) acoustic phonons $R_{\rm TA}(R_{\rm LA})$ versus the ratio 
$\omega_a/\omega_{\rm B}$. (d) The magnetoresistance induced by intravalley homopolar (Fr\"ohlich coupling) optical phonons 
$R_{\rm HP}(R_{\rm FR})$ versus the ratio $\omega_o/\omega_{\rm B}$ at temperatures $T=100,200,300\,{\rm K}$. 
The other parameters are the same as Fig.\,\ref{SdHT}.}\label{mpr}
\end{figure}

Now we concentrate on the case of higher temperature up to room temperature. First, we consider the suspended MoS$_2$. 
The total magnetoresistances $R_{xx}$ induced by the intravalley screened Coulombic electron-impurity scattering ($d=0$) 
and all above-mentioned intravalley and intervalley electron-phonon couplings except SOP mode, 
are plotted as functions of magnetic field at various high temperatures in Fig.\,\ref{mpr}(a). 
The $R_{xx}$ increases with the increment of magnetic field, accompanying an oscillation at large fields. 
The behavior of resistivity increase with increasing magnetic field is due to impurity-induced resistivity $R_{\rm im}$ as shown 
in Fig.\,\ref{mpr}(b). Since SdHO almost disappears at this temperature, the small oscillation in $R_{xx}$ originates 
from phonon scatterings. With ascending temperature, $R_{\rm im}$ descends, while the total $R_{xx}$ increases 
because of the increasing contributions from electron-phonon scatterings. It is found that, in addition to the electron-impurity scattering, 
the contributions of the intravalley transverse and longitudinal acoustic phonons, $R_{\rm TA}$ and $R_{\rm LA}$, and those of
homopolar and Fr\"ohlich coupling optical phonons, $R_{\rm HP}$ and $R_{\rm FR}$,  play a dominant role in the total resistivity.  
The inset of Fig.\,\ref{mpr}(a) shows that the oscillation arises mainly from the optical contribution $R_{\rm op}=R_{\rm HP}+R_{\rm FR}$. 
The acoustic one $R_{\rm ac}=R_{\rm TA}+R_{\rm LA}$ gives almost a constant value at room temperature.

Actually, the acoustic contributions $R_{\rm TA}$, $R_{\rm LA}$ also oscillate with magnetic field, especially at relatively low temperature, 
exhibiting the so called MPR induced by acoustic phonons.\cite{zudov2001new, zhang2008resonant, lei2008low, wang2013nonlinearm} 
The magnetoresistance peaks occurs when the energy of the optimum phonons $\omega_a=2k_{\rm F}v_{\rm ac}$ equals an integral multiple 
of the inter-Landau-level distance $\omega_{\rm B}$ near Fermi surface. Here $v_{\rm ac}=v_{\rm TA}$ or $v_{\rm LA}$ is the sound velocity 
for the transverse or longitudinal mode. The energy distance $\omega_{\rm B}$ between two intravalley Landau levels with same spin  
around Fermi energy $\varepsilon_{\rm F}$ is given by 
\begin{equation}
\omega_{\rm B}\approx\frac{\omega_c^2}{2\sqrt{(\bar\Delta-\tau s\bar\lambda)^2+\nu_{\tau s}\omega_c^2}}\approx\frac{\omega_c^2}{\Delta}.
\end{equation}
Fig.\,\ref{mpr}(c) indeed shows the oscillation of magnetoresistance for both $R_{\rm TA}$ and $R_{\rm LA}$ with inverse magnetic field 
having period $\varDelta(\omega_a/\omega_{\rm B})\simeq 1$. With increasing temperature, the peaks at high ratio $\omega_a/\omega_{\rm B}$ 
tend to disappear gradually. Further, the peak slightly shifts to smaller $\omega_a/\omega_{\rm B}$ position and the magnetoresistance 
due to longitudinal mode $R_{\rm LA}$ becomes larger than the transverse one $R_{\rm TA}$ in view of enlarged phonon energy 
with the rise of temperature. In the present case with relatively low mobility, the MPR induced by acoustic phonons 
has little influence on the total magnetoresistance. However, for monolayer MoS$_2$ having ultrahigh mobility, the acoustic electron-phonon 
coupling contributes dominantly at low temperature, hence this MPR could be observable.

\begin{figure}
\begin{center}
  \includegraphics[width=0.48\textwidth]{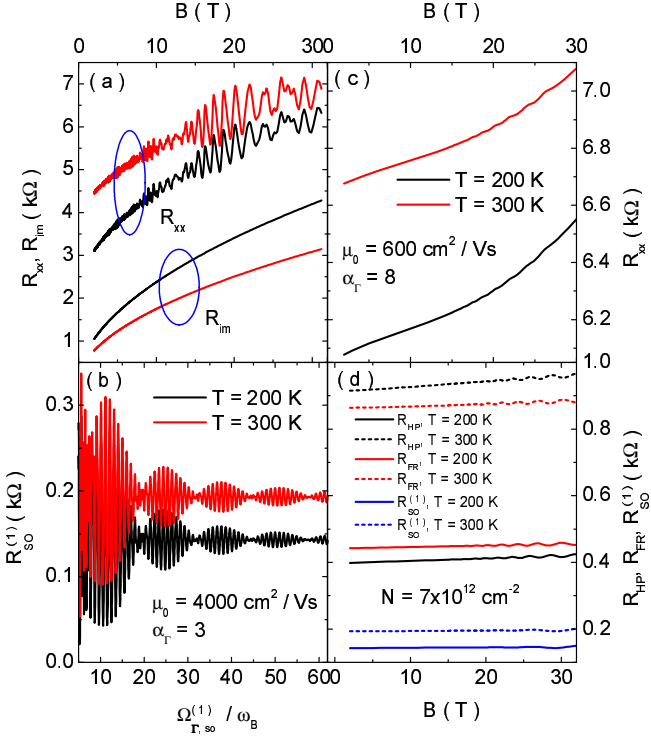}
\end{center}
\caption{(Color online) Magnetoresistance in ZrO$_2$/MoS$_2$/Air structure at high temperature $T=200,300\,{\rm K}$. 
Here the electron density $N=7\times10^{12}\,{\rm cm^{-2}}$. In (a) and (b), the zero-field mobility at zero temperature 
$\mu_0=4000\,{\rm cm^2/Vs}$ and $\alpha_{\Gamma}=3$, while in (c) and (d) $\mu_0=600\,{\rm cm^2/Vs}$ and $\alpha_{\Gamma}=8$ 
for another sample. (a) Total magnetoresistance $R_{xx}$ and impurity-induced one $R_{\rm im}$ versus magnetic field $B$. 
(b) The SOP-induced magnetoresistance $R_{\rm SO}^{(1)}$ is plotted as a function of $\Omega_{\bf \Gamma,\rm so}^{(1)}/\omega_{\rm B}$. 
(c) The total magnetoresistance for the sample with low mobility versus the magnetic field. 
(d) The magnetoresistance induced by intravalley homopolar, Fr\"ohlich coupling optical phonons, and SOP $R_{\rm HP},R_{\rm FR}$, 
and $R_{\rm SO}^{(1)}$ versus magnetic field.}\label{mprsop}
\end{figure}

With further increase in lattice temperature, the electron-optic phonon coupling becomes more and more important in comparison with
other scattering mechanisms. Due to the large coupling coefficients, the resistivities induced by the homopolar and Fr\"ohlich interactions
have largest values.  It is well known that the resistivity exhibits MPR when the energy of optical phonons $\omega_o$ 
equals the distance of Landau levels. In monolayer MoS$_2$ for $\varepsilon_{\rm F}-\bar\Delta\ll\bar\Delta$, the intravalley Landau levels 
are almost evenly spaced and the level distance approximately equals $\omega_{\rm B}$. In Fig.\,\ref{mpr}\,(d) the magnetoresistance $R_{\rm HP}$ or $R_{\rm FR}$ 
is plotted versus the ratio $\omega_o/\omega_{\rm B}$, 
where $\omega_o=\Omega_{\bf \Gamma,{\rm HP}}$ or $\Omega_{\bf \Gamma,{\rm LO}}$ is respectively the frequency for homopolar or Fr\"ohlich coupling. 
It is true that the magnetoresistances show peaks or valleys at $\omega_o/\omega_{\rm B}=l$, i.e. magnetoresistance oscillates 
with inverse magnetic field having period $\varDelta(\omega_o/\omega_{\rm B})\simeq1$. 
However, in contrast to the usual MPR induced by optical phonons in two-dimensional electron gas, 
the oscillating resistivity in MoS$_2$ is modulated due to the spin splitting by an approximate factor 
$\cos\left(2\pi\frac{\omega_o}{\omega_{\rm B}}\frac{\lambda}{\varepsilon_{\rm F}+\bar\Delta}\right)$ analogous to the SdHO. 
Hence, there are nodes at $\frac{\omega_o}{\omega_{\rm B}}\frac{\lambda}{\varepsilon_{\rm F}+\bar\Delta}=l\pm\frac{1}{4}$. 
This leads to the nodes appearing at ${\omega_o}/{\omega_{\rm B}}=19.1,31.9$, in accordance with Fig.\,\ref{mpr}(d).

 Now we study the MPR for MoS$_2$ on a ZrO$_2$ substrate. It is found that the frequencies of SOPs for ZrO$_2$ 
are so small that they play an important role in electron transport.\cite{Macharge2014} Hence, in Fig.\,\ref{mprsop} magnetoresistances 
for MoS$_2$ on ZrO$_2$ are plotted versus magnetic field. In the calculation, the elastic scattering is assumed to be the intravalley 
remote-impurity scattering distributing at $d=1\,$nm from the single layer, the inelastic scatterings are due to all the intrinsic modes 
mentioned above and the intravalley SOPs. $q_{\rm TF}^{\rm eff}$ used in the screening of elastic scattering is estimated to be $0.3q_{\rm TF}$ 
from Fig.\,2 in Ref.\,\onlinecite{Macharge2014}. 
Two samples with different zero-field mobilities are considered for comparison. 
For the clean system, the remote impurity scattering weakens the quick increase of impurity-induced resistivity with magnetic field 
in contrast to suspended case, leading to more evident MPR in the total magnetoresistance. On the other hand, in comparison to suspended MoS$_2$, 
the MPR behavior becomes more complex because of the crucial influence of SOPs, especially the mode with low frequency 
$\Omega_{\bf\Gamma,\rm so}^{(1)}$. The resistivity $R_{\rm SO}^{(1)}$ induced by the first SOP mode is plotted in Fig.\,\ref{mprsop}(b) 
versus $\Omega_{\bf\Gamma,\rm so}^{(1)}/\omega_{\rm B}$. The resonant feature of $R_{\rm SO}^{(1)}$ is similar to other intrinsic modes. 
However, for another sample with low mobility in heavily overlapping-Landau-level regime, the MPR almost disappear. 
The magnetoresistance increases monotonously with magnetic field, and only a small oscillation occurs at very large field. 
In Fig.\,\ref{mprsop}(d), contributions of three important optical modes are plotted.

 The effect of a SiO$_2$ substrate on the MPR of MoS$_2$ is also tested 
and it is found that this dielectric plays negligible role due to its large frequencies of SOPs.

\section{Summary}

In summary, we have studied the linear magnetotransport in single layer MoS$_2$ employing a balance equation analysis by including 
spin-orbit coupling and all kinds of intravalley and intervalley electron-impurity and electron-phonon scatterings.

The existence of an energy gap between the conduction and valence bands, or lack of electron-hole symmetry of the zero Landau level in MoS$_2$, 
makes its magnetotransport behavior more like a conventional 2D electron gas than graphene: 
the resistivity peaks or valleys of its low-temperature SdHO, resulting either from intravalley or from intervally elastic scatterings,
locate at integers of filling factor $\nu_0$.

The large spin-orbit coupling in the system, however, gives rise to a significant modulation or beating of the magnetoresistance oscillation,
or a phase inversion of the oscillation peaks.  
The agreement between theoretical prediction and recent experiment on the phase inversion of SdHO peaks 
demonstrates the importance of the spin-orbit splitting in magnetotransport even for systems of low-mobility. 
 The clear beating pattern of the oscillating magnetoresistance should appear in the well-separated Landau-level regime in high-mobility systems.

On the other hand, the behavior of magnetoresistance oscillation at large magnetic fields or small filling factors appears different 
for intravalley and intervalley scatterings: the period of oscillation associated with intravalley scattering may halve 
due to the weak decay of the second-order oscillating term, while in the case of intervalley disorder much stronger spin-orbit induced 
SdHO modulation shows up that there exists a magnetic-field range in which the magnetoresistivity almost vanishes. 
Of course, intervalley elastic scattering contributes only a much smaller part to the total magnetoresistance than that 
from intravalley ones.

At high temperatures, the magnetoresistance oscillation arising from MPR may show up 
in the smooth impurity-induced resistivity background both for suspended and nonsuspended samples with high mobility. 
Both acoustic phonons (mainly intravalley transverse and longitudinal acoustic modes) and optic phonons 
(mainly homopolar and Fr\"ohlich modes) can induce MPR. 
A beating pattern with the same frequency as in the SdHO also appears in the optical-phonon-induced MPR due to spin-orbit coupling.
For the single layer on a substrate, another resonance due to SOPs 
may occur, resulting in a complex behavior of the total magnetoresistance. 
However, for nonsuspended layer with low mobility, the magnetoresistance oscillation almost disappears 
and the resistivity increases with field monotonously.

\section*{ACKNOWLEDGMENTS}
This work was supported by the National Basic Research Program of
China (Grant No. 2012CB927403) and the National Science Foundation of China
(Grant No. 11474005).


\end{document}